\documentclass[12pt,onecolumn]{article}
\usepackage{amssymb,amsmath,epsfig,graphics,graphicx,enumerate,float,subcaption,verbatim,xspace, color,url}
\usepackage{multirow}
\usepackage{epstopdf}

\newcommand{\nop}[1]{}
\newcommand{\ie}{{\sl i.e.}\xspace}

\begin{document}

\title{A Mixture Model Based Defense for Data Poisoning Attacks Against Naive Bayes Spam Filters}

\author{
David~J.~Miller, Xinyi Hu, Zhen Xiang, and George Kesidis\thanks{This research is 
supported in part by an AFOSR DDDAS grant
and a  Cisco Systems URP gift.}\\
~\\
School of EECS\\
The Pennsylvania State University\\
University Park, PA 16802\\
\{djm25, xzh106, zux49, gik2\}@psu.edu
}

\maketitle

\begin{abstract}
Naive Bayes spam filters are highly susceptible to data
poisoning attacks.  Here, known spam sources/blacklisted IPs exploit the fact that
their received emails will be treated as (ground truth) labeled spam examples, and used for classifier training (or re-training).  The attacking source thus generates emails that will skew the
spam model, potentially resulting in great degradation in classifier accuracy.   Such attacks are successful
mainly because of the poor representation power of the naive Bayes (NB) model, with only a single (component) density to represent
spam (plus a possible attack).  We propose a defense based on the use of a {\it mixture} of NB models.  We demonstrate that the learned mixture almost completely isolates the attack in a second NB component, with the original spam component essentially unchanged by the attack.
Our approach addresses both the scenario where the classifier is being re-trained in light of new data
{\it and}, significantly, the more challenging scenario where the attack is embedded in the original spam training set.
Even for weak attack strengths, BIC-based model order selection chooses a two-component solution, which invokes the mixture-based defense.  Promising results are presented on the TREC 2005 spam corpus.

\end{abstract}

\section{Introduction}
Interest in adversarial learning has grown dramatically in recent years, with some works focused on devising attacks
against machine learning systems,
e.g., \cite{szegedy2014,papernot2016}, and others devising 
defenses, e.g., \cite{meng2017,MLSP18-ADA,MLSP17}.
In this work, we address data poisoning attacks on generative classifiers, with particular focus on naive
Bayes spam filters.  Recent reviews  of spam filtering are 
\cite{Laorden14,ADC-spam,cormack2008};   see also
\cite{trigger-words12,trigger-words13}.
In data poisoning against spam filtering, the attacker, using an IP address known to produce spam, generates emails that
will be ground-truth labeled as spam but which are more representative of ham.  If these emails, and in sufficient quantity, are added to the spam training set, they will
grossly alter the word distribution under the spam model, with concomitant degradation in
accuracy of the spam filter \cite{Tygar11}.

Defenses against data poisoning attacks on various systems include 
\cite{steinhardt2017} and works described in \cite{wild}.
For spam filtering, a ``data sanitization'' strategy was proposed \cite{RONI}, which rejects putative additional training samples if trial-adaptation of the spam model based on use of these
samples causes degradation in classification accuracy on a held-out validation set.  This strategy
makes two assumptions: i) that there is sufficient labeled data to have a held-out validation
set; ii) it assumes the classifier has already been trained on ``clean'' data, with the attack consisting
of additional labeled samples for classifier {\it retraining}.  \cite{RONI} is not a practical
strategy when the attack is {\it embedded} within the original training set (with the attack samples an unknown subset) --- in such a case,
a strategy such as \cite{RONI} would entail an enormous combinatoric space of $\{$clean spam subset, attack data subset$\}$ hypotheses to explore.

A {\it potential} strategy for designing a system robust to an {\it embedded} data poisoning attack is to identify the attack samples as training set outliers.
While such ideas are mentioned in \cite{Tygar11} and are related to \cite{Miller_Browning}, we are not aware such ideas have been practically, effectively applied to spam filtering.  The reason naive
Bayes spam filters are so susceptible to data poisoning attacks is because, under each class hypothesis
(ham or spam) there is only a {\it single} model, whose training/model estimation is degraded, in an
unimpeded fashion, by the attacker. We propose a model and learning mechanism that effects isolation
of the attack samples, so that they have little effect on estimation of the learned spam word model.  Instead of modeling spam using a single NB word distribution, we propose a {\it mixture} of NB models for spam.  We couple this mixture model with both
i) a careful component initialization strategy, so that the second component captures and isolates the attack and ii) BIC-based model selection, which determines whether a second mixture component is warranted
(and hence whether the defense is invoked).  Experimentally, whenever the attack strength is sufficient to induce even the mildest degradation in classification accuracy, BIC selects a two-component model, which invokes the defense and its attack mitigation.  Moreover, most importantly, we emphasize that our defense addresses {\it both} the scenarios where the attack is on classifier {\it retraining} and when the attack is on the {\it initial} training. The latter is the more challenging scenario.
\section{A Mixture-based Defense Against Poisoning of Spam Filters}
\subsection{Notation}
We consider a dictionary of $N$ unique words (following standard stemming and stoplisting), with a given email represented as a vector of word counts
$\underline{x} = (x_1,x_2,\ldots,x_N)$, $x_l$ the number of times word $l$ from the dictionary
occurs in the given email.  Thus, emails are represented by fixed high-dimensional (but highly sparse)
vectors.  Let ${\cal X}_h = \{\underline{x}_i^{h}, i=1,\ldots, T_h\}$ be a given training set of
ham emails, used to build a NB ham model.  Likewise, let ${\cal X}_s = \{\underline{x}_i^{s}, i=1,\ldots, T_s\}$ be a training set of spam emails for building a NB spam model.
\subsection{Attack Scenarios}
We consider two attack scenarios: i) classifier retraining
and ii) classifier {\it training}, which is more challenging, and for which our method is the first successful defense of which we are aware.  In the 
first (retraining) scenario, one can initially build {\it clean} ham and spam models (those uncorrupted
by attack) using ${\cal X}_h$ and ${\cal X}_s$, respectively\footnote{Note that maximum likelihood estimation or Bayesian variants are easily performed, using normalized frequency count estimates (over the labeled
corpus) for each word, conditioned on the class.}.
Let us denote a batch of additional samples that are treated as labeled spam by
${\cal \tilde{X}}_s = \{\underline{\tilde{x}}_i,i=1,\ldots,\tilde{T}_s\}$.  In the  
retraining case, the learner pools ${\cal \tilde{X}}_s$ with ${\cal X}_s$, retraining the
spam model using the combined data pool ${\cal X}_{sc} = \{{\cal X}_s,{\cal \tilde{X}}_s\}$.
Note that ${\cal \tilde{X}}_s$ may consist of legitimate spam samples, attack samples,
or some combination of the two.  If one can utilize a separate, uncorrupted, held-out
validation set, the approach in \cite{RONI} can effectively mitigate an attacking ${\cal \tilde{X}}_s$.  However, consider the other scenario -- the {\it training} scenario.  Unlike retraining, where
the subset ${\cal \tilde{X}}_s$ is known to the learner, in the training 
scenario the attack samples are embedded amongst the clean spam samples.  The learner does not know whether an attack is present and if so, which is the attack sample subset.  Again the learner uses
 ${\cal X}_{sc}$, but in this case to perform the inaugural learning of the spam model, not model 
retraining.  In the sequel, we develop a common mixture-based defense strategy, which effectively defeats
the attack in {\it both} scenarios. 
\subsection{Two-component Mixture Model for Spam}
The standard NB spam classifier computes the maximum {\it a posteriori} (MAP) decision,
given an email's vector $\underline{x}$, as $\hat{c}_{\rm MAP}(\underline{x}) = {\rm arg max}_{c} P[C=c | \underline{x}] = {\rm arg max}_c \log(\alpha_c) + \sum\limits_{i=1}^N x_i \log(P[x_i | c])$,
where $c \in \{h,s\}$, $\alpha_s + \alpha_h =1$, and $P[x_i | c]$ is the probability of count $x_i$ for word $i$ under (a multinomial model for) class $c$,
i.e. there is a single NB component model for each class.
We will invoke a BIC-based hypothesis testing strategy \cite{Schwarz} to decide, given ${\cal X}_{sc}$,
whether to use a single or a two-component model for spam\footnote{Our approach can be extended to evaluate models with more than two components if there is more than one attacker, a single but multi-modal attack, or if there is both attack {\it and} legitimate class drift reflected in  ${\cal X}_{sc}$.}.
The two-component spam model is 
\begin{eqnarray*}
P_M[\underline{x} | s] & = & 
 \beta_1 \prod\limits_{i=1}^N P[x_i | C=s,M=1]\\
& & ~~ + \beta_2 \prod\limits_{i=1}^N P[x_i | C=s,M=2],
\end{eqnarray*}
where $\beta_1+\beta_2=1$, $M$ the mixture component random variable.
In principle, the class MAP rule can be invoked as before, evaluating for the spam case
$\log(\alpha_s) + \log(P[\underline{x} | s)$.  However, in the spam case, due to the two components, one no longer gets a sum of logs expression and numerical underflow practically inviolates evaluation of $\log(P[\underline{x} | s)$.

However, this can be mitigated without any approximation of the MAP decision rule, as follows.
Note that 
$$P[\underline{x} | C=s,M=j] = \frac{P[M=j,C=s | \underline{x}]P[\underline{x}]}{P[C=s,M=j]}, j=1,2.$$
Then, substituting
this expression into 
\begin{eqnarray*}
\log(P[\underline{x} |s]) & = & \log(\beta_1 P[\underline{x} | C=s, M=1]\\
& & ~~ + \beta_2 P[\underline{x} | C=s, M=2]),
\end{eqnarray*}
we obtain 
$$\log(P[C=s,M=1 | \underline{x}] + P[M=2,C=s | \underline{x}]) + \log(P[\underline{x}]).$$
Similarly, the MAP decision statistic under ham can be expressed as 
$$\log(P[C=h | \underline{x}]) + \log(P[\underline{x}]).$$
The term $\log(P[\underline{x}])$ is common under the ham and spam expressions, and can be ignored.
Finally, we note that the posterior probabilities $P[C=h | \underline{x}]$, $P[C=s,M=1 | \underline{x}]$, and $P[M=2,C=s | \underline{x}]$ are easily computed, and without numerical underflow problems.  Thus, exact ham vs. spam MAP inference in the two-component spam case is readily achieved.
\subsection{EM Learning of the Spam Mixture} 
Maximum likelihood estimation of the two-component mixture can be performed via a standard application of the
Expectation-Maximization (EM) algorithm \cite{Dempster}.  The model parameters are $\{\beta_1,\beta_2, \{\lambda_{l|C=s,M=1}\},
\{\lambda_{l | C=s,M=2}\}\}$, the $\{\lambda\}$ the multinomial word probabilities under the two components.

\noindent
{\it E-Step:}
\newline
Given the current model parameters\footnote{The first time the E-step is invoked, it uses the initialized model parameters.}, one computes 
$$P[M=j | \underline{x}, C=s] = \frac{\beta_j \prod\limits_{i=1}^N \lambda_{i|s,j}^{x_i}}{\sum\limits_{j'=1,2} \beta_{j'} \prod\limits_{i=1}^N \lambda_{i|s,j'}^{x_i}} \forall \underline{x} \in {\cal X}_{sc}.$$

\noindent
{\it M-Step:}
\newline
The parameters are re-estimated via:
\begin{eqnarray}
\beta_j =  \frac{\sum\limits_{\underline{x} \in {\cal X}_{sc}} P[M=j | \underline{x},s]}
{T_s + \tilde{T}_s}, j=1,2
\end{eqnarray}
\begin{eqnarray}
\lambda_{i|s,j} = \frac{\sum\limits_{\underline{x} \in {\cal X}_{sc}} x_i P[M=j | \underline{x},s]}
{\sum\limits_{\underline{x} \in {\cal X}_{sc}} N_{\underline{x}} P[M=j | \underline{x},s]}, \forall i \end{eqnarray}
where $N_{\underline{x}}$ is the number of words in the given document\footnote{We slightly modify these updates, giving $\epsilon =10^{-6}$ additional counts to all words, to ensure
that all words have non-zero probabilities.}.
The E and M steps are alternated until a convergence criterion is met, with these iterations strictly increasing in the
log-likelihood of ${\cal X}_{sc}$.  The same EM algorithm applies in both the training
and retraining cases, except for one key difference -- the model parameter initialization, next discussed.

After model learning, the component whose emails are MAP-assigned with a large percentage to the ham model (ham versus component model) is discarded. This removes the attack component. The remaining component is a nearly `pure spam' component.
 
\subsection{Spam Model Initialization}
Crucial to our defense is the choice of initial model parameters, seeding the EM learning.
\newline
{\it Retraining Scenario:}

\noindent
In this case, since we have already learned the inaugural (single component) ham and spam models, we can compute the ham posterior probability for $\underline{x}\in{\cal \tilde{X}}_s$: 
$$P[C=h | \underline{x}] = \frac{\alpha_h P[\underline{x} | C=h]}{\alpha_h P[\underline{x} | C=h] + \alpha_s P[\underline{x} | C=s]}, \forall \underline{x} \in{\cal \tilde{X}}_s.$$  
Since an attack introduces emails labeled as spam but with characteristics of ham, we suggest
to initialize the multinomial word probabilities as follows:
\begin{eqnarray}
\lambda_{l | M=2, C=s} = \frac{\sum\limits_{\underline{x} \in {\cal \tilde{X}}_s} x_l P[C=h |\underline{x}]}{\sum\limits_{\underline{x} \in {\cal \tilde{X}}_s} N_{\underline{x}} P[C=h | \underline{x}]}, \forall l
\end{eqnarray}
\begin{eqnarray}
\lambda_{l | M=1, C=s} = \frac{{\ \text{count of word}} \ l \ \text{in} \ {\cal{X}}_s}{{\ \text{count of words in}} \ {\cal{X}}_s}, \forall l.
\end{eqnarray}
Again, $\epsilon$ counts are added to avoid zero probabilities. Here, component 1 is seeded to be the {\it true} spam component, with component 2 seeded to capture the attack. We initialize $\beta_1 = \beta_2 = 0.5$.

\noindent
{\it Training Scenario:}

In this case, the learning of the two components is performed on the combined data pool ${\cal X}_{sc}$. The multinomial word probabilities are initialized as follows:
\begin{eqnarray}
\lambda_{l | M=1, C=s} = \frac{{\ \text{count of word}} \ l \ \text{in} \ {\cal{X}}_{sc}}{{\ \text{count of words in}} \ {\cal{X}}_{sc}}, \forall l
\end{eqnarray}
\begin{eqnarray}
\lambda_{l | M=2, C=s} = \frac{{\ \text{count of word}} \ l \ \text{in} \ {\cal{X}}_h}{{\ \text{count of words in}} \ {\cal{X}}_h}, \forall l.
\end{eqnarray}

Note that the ``attack'' component is initialized using the ham data. We realize that the initial esitimation of $\lambda_{l | M=1, C=s}$ is contaminated by the attack. However, the subsequent EM algorithm is effective at  ``undoing'' this contamination,  \ie at reassigning the attack emails to component 2. 
Again, $\beta_1 = \beta_2 = 0.5$, is used for initialization of EM.

\section{Experimental Results}

We used a subset of the TREC 2005 spam corpus \cite{TREC05}, 
also used in \cite{Tygar11}.  The training sets consisted of 8000 ham
and 7977 spam emails.  The (exclusive) test set consisted of 2000 ham and 1994 spam emails. The dictionary
(following stemming and stoplisting) consisted of 19080 words.  Attack emails were generated
as follows: i) the email length was randomly chosen to be the length of one of the spam training emails; ii)  Then the words were generated i.i.d. based on the NB ham model.  
We also evaluated a second attack distribution which only used the words {\it{more likely}} under ham than spam -- this truncation of the dictionary (to 8645 words) makes the attack more potent.
The test set accuracy is evaluated for different attack strengths varying from 0 to $10^5$ attacking emails in each scenario.

{\it Retraining Scenario:}
\begin{figure}[h]
\begin{center}
\includegraphics[width=4.5in]{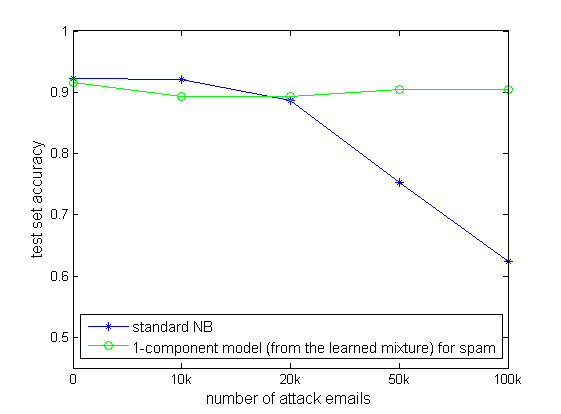}
\caption{Retraining-- test set accuracy with `pure-ham' attacks.}
\label{fig:retrain_pureham}  
\end{center}     
\end{figure}

\begin{figure}[h]
\begin{center}
\includegraphics[width=4.5in]{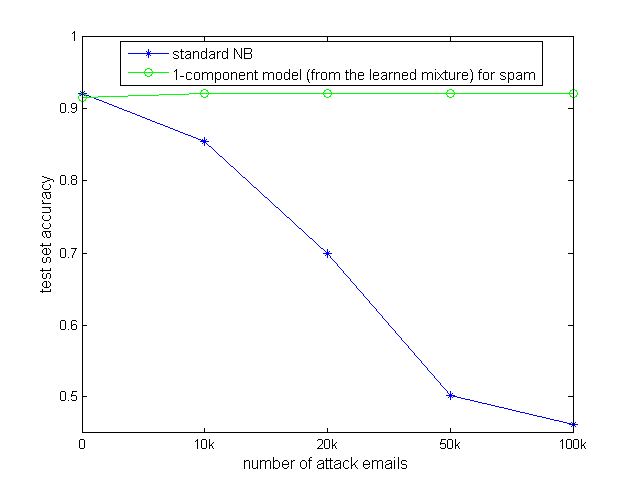}
\caption{Retraining-- test set accuracy with `truncated' attacks.}
\label{fig:retrain_hamdiff}  
\end{center}     
\end{figure}

For the retraining scenario, our method is compared with the standard NB method for both attacks. As shown in Fig. \ref{fig:retrain_pureham}, under the `pure-ham' attack, the test accuracy for the standard NB drops rapidly as the attack strength grows over $2\times 10^4$. Our method, however, keeps at a high test accuracy (around 0.9) for all attack strengths. For the `truncated' attack case where the attack is more potent, the test accuracy of the standard NB drops even faster. Our method still performs well, showing its robustness to variation in the attack strength.

In addition to the results shown in the figures, we also address two points. First, we compared the BIC values between the single-component model and the two-component model. As the number of attack emails (except zero) are tested, the two-component model is always preferred, with a lower BIC. Second, the learned mixture almost completely isolates the attack in the second NB component, and the original spam component is essentially unchanged by the attacks. Especially, none of the attacks are classified as true spam in the mixture in all cases for `truncated' attacks above (except $10^4$, for which one attack is misclassified as true spam).

%Fig. \ref{fig:retrain_pureham} shows the test accuracy for the retraining scenario with the pureham attacks, and Fig. \ref{fig:retrain_hamdiff} is for the truncated attacks. The blue line is the result of standard NB with one component for each class, and the green line is the result of 1-component model (from the learned mixture) for spam which ignores the attack’s cluster of the learned mixture. Results show BIC-based model selection chooses a two-component solution except attack num$=0$. In the truncated attacks' experiments, we also observed  the Kullback–Leibler distance between the 1-component model for spam which comes from the learned mixture and the pure ham model is almost unchanged with different number of attacks. The true spam component is not affacted by attacks. 

{\it Training Scenario:}
\begin{figure}[h]
\begin{center}
\includegraphics[width=4.5in]{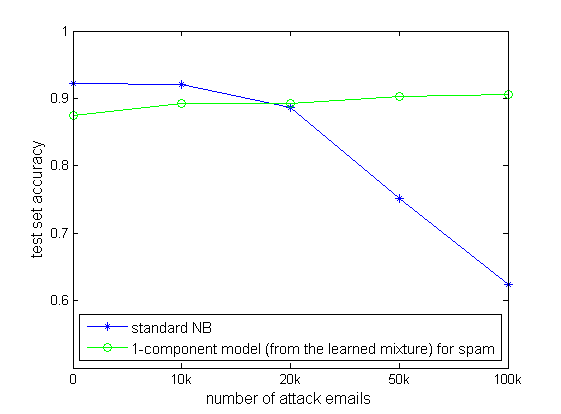}
\caption{Training-- test set accuracy with `pure-ham' attacks.}
\label{fig:train_pureham}  
\end{center}     
\end{figure}

For the training scenario, the spam distribution is unknown a priori. Thus, the `truncated' attack exploits knowledge that is unavailable even to the designer of the classifier. Thus, we do not evaluate this attack in the training case, focusing on the `pure-ham' attack. Fig. \ref{fig:train_pureham} shows that our method has similar performance to the standard NB for low attack strengths and beats the standard NB dramatically when the attack strength is high.

%Fig. \ref{fig:train_pureham} shows the test accuracy for the training scenario with the pureham attacks. Results show BIC-based model selection chooses a two-component solution except attack num$=0$.

\section{Discussion}
We have considered attacks which corrupt the spam data.  Our approach could also be applied if the attack targets ham,
rather than spam.  However, it is more complicated to address an attack that simultaneously poisons both spam and ham.
That is a good subject for future work. Another scenario of interest is where there is both an attack and legitimate ``class drift'',
e.g. a time-varying distribution for spam.  In such a case, one component may be needed to model spam class drift, with another capturing the attack.  It
may be possible to identify these two components because we would
expect the attack distribution to be closer to the ham distribution than legit drifting spam. Another good research direction
is to apply parsimonious mixture modeling \cite{GrahamMiller2006} to learn accurate spam and attack components, working in the full
word space.  This approach is much more computationally complex, but should also be a highly accurate spam model ``initialization'' approach. 
\section{Conclusions}
In this work, we proposed a mixture model based defense against data poisoning attacks against spam filters, devising defenses against
attacks on both classifier re-training as well as against {\it initial} classifier training.  Our approach should be
applicable more generally to defend against data poisoning attacks on other domains that involve generative modeling of the data.
Our mixture-based approach could also be applied as a pre-processor, to remove attacks before they corrupt (discriminative) training of a deep neural network or support vector machine based classifier.

\clearpage
\bibliographystyle{plain}
%\bibliography{./bibfiles/adversarial,./bibfiles/adversarial0,./bibfiles/kesidis-prior,./bibfiles/transductive,./bibfiles/spam}

\end{document}